\documentclass[twocolumn,pre,amsmath,amssymb,showpacs]{revtex4}

\usepackage{amsfonts} %,psfig}
\usepackage[dvips]{graphicx}
\usepackage{epsfig}

\usepackage{amsmath}
\usepackage{amssymb}

\begin{document}
\newtheorem{theorem}{Theorem}
\newtheorem{definition}{Definition}

\topmargin = -0.0in

%%%%%%%%%%%%%%%%%%%%%%%%%%%

\newcommand{\mbf}[1]{\mbox{\boldmath $#1$}}

\def\nequiv{\;{/}{\!\!\!\! }{\equiv}\;}
\def\R{\mbox{$I\!\!R$}}
\def\Z{\mbox{$I\!\!Z$}}
\def\b{\mbox{\mbf b}}
\def\c{\mbox{\mbf c}}
\def\0{\mbox{\mbf 0}}
\def\G{\mbox{\mbf G}}

\def\f{\mbox{\mbf f}}
\def\g{\mbox{\mbf g}}
\def\h{\mbox{\mbf h}}
\def\u{\mbox{\mbf u}}
\def\x{\mbox{\mbf x}}
\def\y{\mbox{\mbf y}}
\def\z{\mbox{\mbf z}}
\def\H{\mbox{\bf H}}

\def\F{\mbox{\bf F}}
\def\s{\mbox{\bf s}}
\def\e{\mbox{\bf e}}
\def\r{\mbox{\bf r}}

\def\1{\mbox{\mbf 1}}

\def\i{\mbox{\mbf i}}
\def\p{\mbox{\mbf p}}
\def\q{\mbox{\mbf q}}

\def\A{\mbox{\mbf A}}
\def\B{\mbox{\mbf B}}
\def\G{\mbox{\mbf G}}
\def\D{\mbox{\mbf D}}
\def\H{\mbox{\mbf H}}
\def\I{\mbox{\mbf I}}
\def\J{\mbox{\mbf J}}
\def\M{\mbox{\mbf M}}

\def\L{\mbox{\mbf L}}
\def\N{\mbox{\mbf N}}

\def\BibTeX{{\rm B\kern-.05em{\sc i\kern-.025em b}\kern-.08em
    T\kern-.1667em\lower.7ex\hbox{E}\kern-.125emX}}

\preprint{APS/123-QED}

\title{Random walks on networks: cumulative distribution of cover time}

\author{Nikola Zlatanov}
\affiliation{
Macedonian Academy for Sciences and Arts, Skopje, Macedonia}
\author{Ljupco Kocarev} 
\affiliation{%
Macedonian Academy for Sciences and Arts, Skopje, Macedonia \\
Institute for Nonlinear Science, University of California, San Diego \\
9500 Gilman Drive, La Jolla, CA 92093-0402}

\date{\today}% It is always \today, today,


\begin{abstract}
We derive an exact closed-form analytical expression for  the distribution  of the cover time for a random walk over an arbitrary graph.  In special case, we derive simplified, exact expressions for the distributions of cover time for a complete graph, a cycle graph, and a path graph. An accurate approximation for the cover time distribution, with computational complexity of $O(2 n)$, is also presented.  The approximation is numerically tested only for graphs with $n\leq 1000$ nodes.  
\end{abstract}

%

%\pacs{05.40.Fb, 02.50.Ga, 02.50.Cw}

%89.75.Hc,02.10.Ox,64.60.ae,64.60.al}

\maketitle

\section{Introduction} 

The random walk is a fundamental dynamic process which can be used to model random processes inherent to many important applications, such as  transport in disordered media \cite{dis}, neuron firing dynamics \cite{neuron}, spreading of diseases \cite{spred} or transport and search processes \cite{search-1,search-2,search-3,search-4,search-5}.

In this paper, we investigate random walks on graphs \cite{laslo} and derive exact expressions for the cumulative distribution functions for three quantities of a random walk that play the most important role in the theory of random walks:
(1) hitting time $h_{ij}$ (or first-passage time), which is the number of steps before node $j$ is visited starting from node $i$; (2) commute time $\kappa_{ij} = h_{ij} + h_{ji}$; and cover time, that is the number of steps to reach every node.

Average hitting time, average commute time, and average cover time have been recently studied in several papers. In \cite{noh} the authors investigate random walks on complex networks and derive an exact expression for the mean first-passage time between two nodes. For each node the random walk centrality is introduced, which determines the relative speed by which a node can receive and spread information over the network in a random process. 
Using both numerical simulations and scaling arguments, the behavior of a random walker on a one-dimensional small-world network is studied in \cite{almas}. The average number of distinct sites visited by the random walker, the mean-square displacement of the walker, and the distribution of first-return times obey a characteristic scaling form. 
The expected time for a random walk to traverse between two arbitrary sites of the Erdos-Renyi random graph is studied in \cite{sood}. 
The properties of random walks on complex trees are studied in \cite{pastor}.  Both the vertex discovery rate and the mean topological displacement from the origin present a considerable slowing down in the tree case. Moreover, the mean first passage time displays a logarithmic degree dependence, in contrast to the inverse degree shape exhibited in looped networks \cite{pastor}. The random walk on networks has also much relevance to algorithmic applications. The expected time taken to visit every vertex of connected graphs has recently been extensively studied. In a series of papers, Cooper and Frieze have studied the average cover time of various models of a random graph, see for example \cite{cover}.

This is an outline of the paper. In section \ref{sec-anal-res} we derive closed formulas of the cumulative distribution function for hitting time, commute time, and cover time; we also present a simple example of a graph with four nodes, and derive closed formulas of the  cumulative distribution function for cover time of complete graphs, cycle and path graphs. An approximation of the  cumulative distribution function for cover time is proposed in the section \ref{sec-approx}; we also present some numerical results of the cumulative distribution function for cover time of different graphs \ref{sec-num}. We finish the paper with conclusions.

\section{Exact Random Walk Distributions for hitting time, commute time, and cover time} \label{sec-anal-res}

Let $G = (V, E)$ be a connected graph with $n$ nodes and $m$ edges. Consider
a random walk on $G$: we start at a node $v_0$; if at the $t$-th step we are at
a node $v_t$, we move to neighbor of $v_t$ with probability $1/d(v_t)$, if an edge exists between node $v_t$ and it's neighbor, where $d(v_t)$ is the degree of the node $v_t$. Clearly, the sequence of random nodes $(v_t: t =0, 1, \ldots)$ is a Markov chain. 
We denote by $M =  (m_{ij})_{i,j \in V}$ the matrix of transition probabilities of this Markov chain: 
\begin{equation}
m_{ij} = \left\{
\begin{array} {cccc}
1/d(i),  & \mbox{if } ij\in E   \\
0, & \mbox{otherwise,} 
\end{array}
\right.
\end{equation} 
where $d(i)$ is the degree of the node $i$. Recall that the probability $m^t_{ij}$ of the event that starting at $i$, the random walk will be at node $j$ after $t$ steps, is an entry of the matrix $M^t$. It is well known that $m_{ij}^t \rightarrow d(j)/2m$ as $t \rightarrow \infty$. 

We now introduce three quantities of a random walk that play the
most important role in the theory of random walks: (1) hitting time $h_{ij}$ is the number of steps before node $j$ is first visited  starting from node $i$; (2) commute time $\kappa_{ij} = h_{ij} + h_{ji}$ is the number of steps in a random walk starting at $i$ before node $j$ is visited and then node $i$ is reached again; and (3) cover time is the number of steps to reach every node. 
%If no starting node (starting distribution) is specified we mean the worst case (i.e. the node from which the cover time is maximum). 

\subsection{Hitting time}

We first calculate the probability mass function for the hitting time. 
To calculate the hitting time from $i$ to $j$, we replace the node $j$ with an absorbing node. Let $D_j$ be a matrix such that $d_{ik} = m_{ik}$ for all $k \neq j$, and $d_{ij}=0$ for all $i\neq j$ and $d_{jj}=1$. This means that the matrix $D_j$ is obtained from $M$ by replacing the original row $j$ with the basis row-vector $e_j$ for which the $j$-th element is 1 and all other elements are 0. 
Let $d_{ij}^t$ be the $ij$ entry of the matrix $D_j^t$, denoting  the probability that starting from $i$ the walker is in the node $j$ by time $t$. Since $j$ is an absorbing state,  $d_{ij}^t$ is the probability  of reaching $j$, originating from $i$, in not more then $t$ steps , i.e. $d_{ij}^t$ is the cumulative distribution function (CDF) of hitting time. Note that the $j-$th column of the matrix $D_j^t$  approaches the all $1$ vector, as $t \rightarrow \infty$. The probability mass function of the hitting time $h_{ij}$ to reach $j$ starting from $i$ is, therefore, given by   
$$
p_{h_{ij}}(t) = d_{ij}^{t} - d_{ij}^{t-1},\quad t\geq 1
$$

Let $E_{x_j}^t$ be the event of reaching the node $x_j$ starting from the node $i \neq x_j$ by time $t$. Consider a sequence of events $\{E_{x_1}^t, E_{x_2}^t, \ldots, E_{x_k}^t\}$. What is the probability of the event: starting from node $i$, the walker visits one of the nodes $\{x_1, x_2, \ldots, x_k\}$ by time $t$? Obviously, it is the probability of the union $\cup _{j=1}^k E_{x_j}^t$.    
To calculate this probability, we replace the nodes $\{x_1,x_2, \ldots, x_k\}$ with absorbing nodes. Let $D_\mathbf{x}$ be a matrix  obtained from $M$ by replacing the rows $\{x_1, x_2, \ldots, x_k\}$  with the basis row-vectors $e_{x_1}, e_{x_2}, \ldots, e_{x_k}$, respectively. 
Let $d_{ix_j}^t$ be the $ix_j$ entry of the matrix $D_\mathbf{x}^t$.   $\sum_{j=1}^k d_{ix_j}^t$ is the probability that starting from $i$ we reach for the first time one of the $\{x_1, x_2, \ldots, x_k\}$ nodes in $\leq t$ steps. Therefore, 
\begin{equation}
F_{x_1, \ldots, x_k}(t) = \sum_{j=1}^k d_{ix_j}^t \label{cumul}
\end{equation}
is the cumulative distribution function (CDF) of the hitting time  $h_{ix}=t$ of the union of events. The probability of reaching one of the nodes $\{x_1, x_2, \ldots, x_n\}$, starting from $i$, in the $t$-th step is given by 
\begin{eqnarray*}
p_{h_{ix}}(t) =   F_{x_1, \ldots, x_k}(t) - F_{x_1, \ldots, x_k}(t-1) ,\quad t\geq 1 
\end{eqnarray*}
which actually gives the probability mass function (PMF) of hitting time $h_{ix}$ of the union  $\cup _{j=1}^k E_{x_j}^t$.

\subsection{Commute time}

Probability mass function  of the commute time $\kappa_{ij} = h_{ij} + h_{ji}$ is obtained as the convolution of PMFs of the two random variables $h_{ij}$ and $h_{ji}$: 
$$
p_{\kappa_{ij}}(t) = p_{h_{ij}}(t) \star p_{h_{ji}}(t)  = \sum_{\tau =1}^{t} p_{h_{ij}}(\tau) p_{h_{ji}}(t - \tau).   
$$
The cumulative distribution function of the commute time
can also be derived as follows: we copy our Markov
chain and we modify the original Markov chain by deleting
all outgoing edges of the node $j$, we modify the original Markov chain by deleting all outgoing edges of the node $j$ and we modify the copied Markov chain by replacing all outgoing edges of the node $i'$ (which is a copy of the node $i$ of the original Markov chain) with a self-loop. We then connect the two chains by adding one directed edge from  node $j$ to its copy $j'$ of the copied chain. Let $O$ be $n \times n$ matrix of all 0s, $O_j = (o_{kl}) $ be the $n\times n$ matrix for which all elements are 0 except $o_{jj}=1$, and $D^*_j$ be the matrix obtained from $M$ by replacing the $j-$th row with all 0. Define the $2n \times 2n$ matrix $C$ as 
$$
C = \left[
\begin{array} {cc}
D^*_j & O_j     \\
O &  D_i   
\end{array}
\right].
$$   
The matrix $C$ is a transition matrix of the modified Markov chain with $2n$ elements (original Markov chain and its copy). Let $c_{i,i+n}^t$ be the $(i,i+n)$ element of the matrix $C^t$. This element is the cumulative distribution function for the commute time $\kappa_{ij}-1$. %The matrix $C$ can also be compute as  
%
%\begin{eqnarray}    
%C &=& I_{2n\times 2n}[i;i+j](I_{2\times 2}\otimes D ) \nonumber \\
%&& \qquad + Z_{2n\times 2n}[(i+j,i+j);(i,i+j)], 
%\end{eqnarray}
%where $\otimes$ is the Kronecker product,  $I_{2\times 2}$ is $2\times 2$ 
%identity matrix, $I_{2n\times 2n}[i;i+j]$ is $2n\times 2n$ 
%identity matrix with $i$-th and $i+j$-th row being zero rows 
%and $Z_{2n\times 2n}[(i+j,i+j);(i,i+j)]$ is a $2\times 2$ 
%zeros matrix with the $(i+j,i+j)$ and $(i,i+j)$ elements being ones.

%

\subsection{Cover time}

Cover time is defined as the number of steps to reach all nodes in the graph. 
%
%What is the probability that the walker visits all nodes by the time $t$ starting from node $z$? 
%
In order to determine the CDF of the cover time, we consider the event   $\cap_{j=1,j\neq z}^n E_{x_j}^t$, and use the well known equation for the inclusion-exclusion of multiple events 
\begin{eqnarray}
P\bigg( \bigcap _{k=1,k\neq z}^n E_{x_k}^t\bigg)= \sum_{i=1,i\neq z}^n P(E_{x_i}^t)-  \nonumber \\  - \sum_{i=1,i\neq z}^n \sum_{j=i+1,j\neq z}^n P(E_{x_i}^t \cup E_{x_j}^t)+ \ldots + \nonumber \\ + (-1)^{n-1} P(E_{x_1}^t \cup E_{x_2}^t \cup \ldots \cup E_{x_n}^t), \nonumber 
\end{eqnarray}
From the last equation and equation (\ref{cumul}), we determine the cumulative distribution function of the cover time as  
\begin{eqnarray}
F_{\mbox{cover}}(t) &=& \sum_{i=1,i\neq z}^n F_{x_i}(t) -    \sum_{i=1,i\neq z}^n \sum_{j=i+1,j\neq z}^n F_{x_i,x_j}(t) +\nonumber \\ 
& &  \ldots + (-1)^{n-1} F_{x_1, x_2, \ldots, x_n}(t).  
  \label{cov-t}
\end{eqnarray}
where $z$ is the starting node of the walk.

Equation (\ref{cov-t}) is the main result of this paper. The probability mass function of the cover time can be easily computed from the Eq.~(\ref{cov-t}).
We note that Eq.~(\ref{cov-t}) is practically applicable only for small values of $n$; in fact the computational complexity of Eq. (\ref{cov-t}) at a single time step is: 
$$
\sum_{j=1}^n\frac{n!}{j! (n-j)!}=2^{n-1}-1. 
$$

\subsection{An Example}

\begin{figure}\center
\resizebox{0.5\textwidth}{!} {\includegraphics{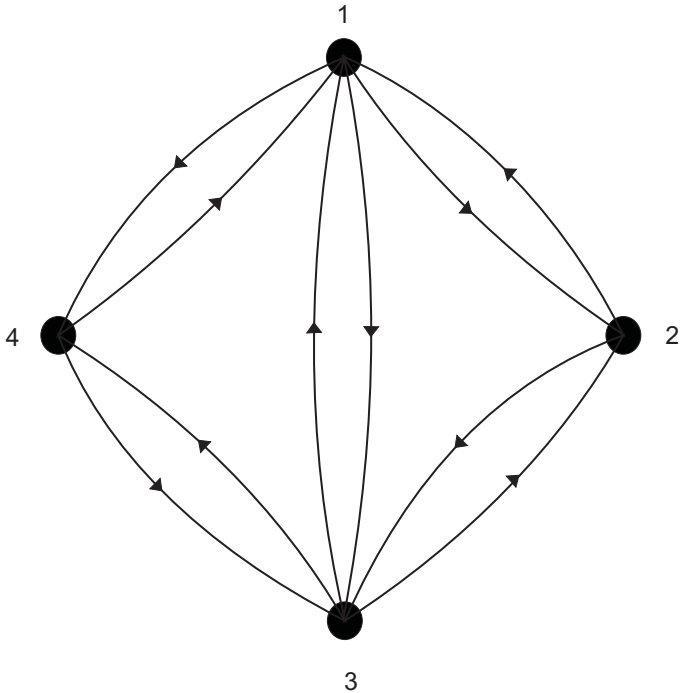}}
\caption{Random walk on a network with four nodes} \label{sl1}\end{figure}

We now present a simple example to illustrate our results. Consider a random walk on a network with four nodes, see Fig.~\ref{sl1}, such that the matrix of transition probabilities of the corresponding Markov chain is given by   
$$
M = \left[
\begin{array} {cccc}
0 & 1/3 & 1/3 & 1/3   \\
1/2 & 0 & 1/2 & 0     \\
1/3 & 1/3 & 0 & 1/3   \\
1/2 & 0 & 1/2 & 0   
\end{array}
\right].
$$
Let $m_{ij}^t$ be the $(i,j)$-th element of the matrix $M^t$. Since, in this example, $M$ is a $4 \times 4$ matrix, one can compute analytically, using for example the software package Mathematica, the elements of the matrix $M^t$. Thus, it can be found, for example,     
$$
m_{14}^t = \frac{1}{5} \left(1 - (-1)^t \left(\frac{2}{3}\right)^t\right), 
$$
which is the probability that the walker starting from $i=1$ at the time $t$ is in $j=4$. Note that $\lim_{t \to \infty} m_{14}^t = 1/5$.

\begin{figure}\center
\resizebox{0.5\textwidth}{!} {\includegraphics{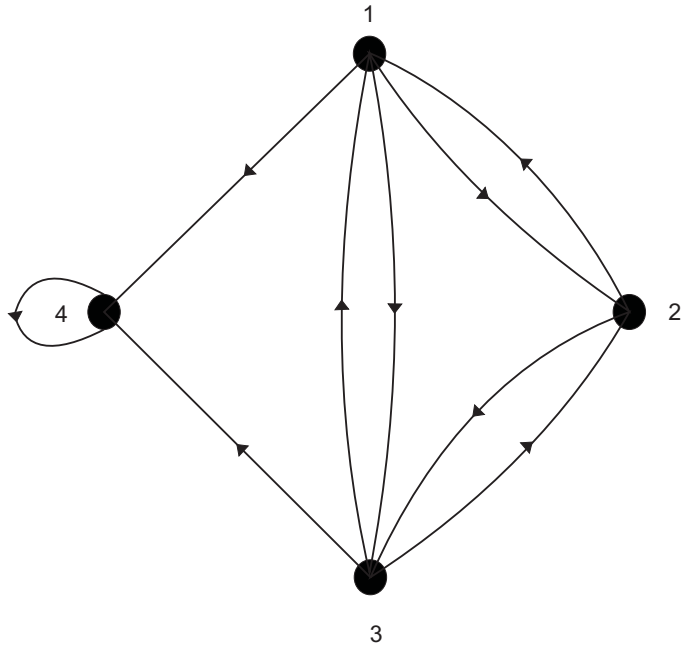}}
\caption{Modified random walk for computing the hitting time to reach the node 4 starting from an arbitrary node} \label{sl2}\end{figure}

To compute the hitting time to reach the node 4 starting from an arbitrary node, we modify the existing random walk to the random walk shown on Fig.~\ref{sl2}. The transition matrix of the modified walk is:  
$$
D_4 = \left[
\begin{array} {cccc}
0 & 1/3 & 1/3 & 1/3   \\
1/2 & 0 & 1/2 & 0     \\
1/3 & 1/3 & 0 & 1/3   \\
0 & 0 & 0 & 1   
\end{array}
\right].
$$
Let $d_{ij}^t$ be the elements of the matrix $D^t$. Again the elements of the matrix  $D^t$ can be computed analytically. For example, the probability of reaching the node 4 starting form 1 in time steps $\leq t$ is equal to 
\begin{eqnarray*}
d_{14}^t = 1 &-& \frac{6^{-t}}{2 \sqrt{13}}\bigg[\left(1-\sqrt{13}\right)^t 
\left(\sqrt{13}-3\right) \nonumber\\&+&\left(1+\sqrt{13}\right)^t 
\left(\sqrt{13}+3\right)\bigg].
\end{eqnarray*}
Clearly, as for any cumulative distribution,  $\lim_{t \to \infty} d_{14}^t = 1$.

\begin{figure}\center
 {\includegraphics[width=9cm,height=8cm]{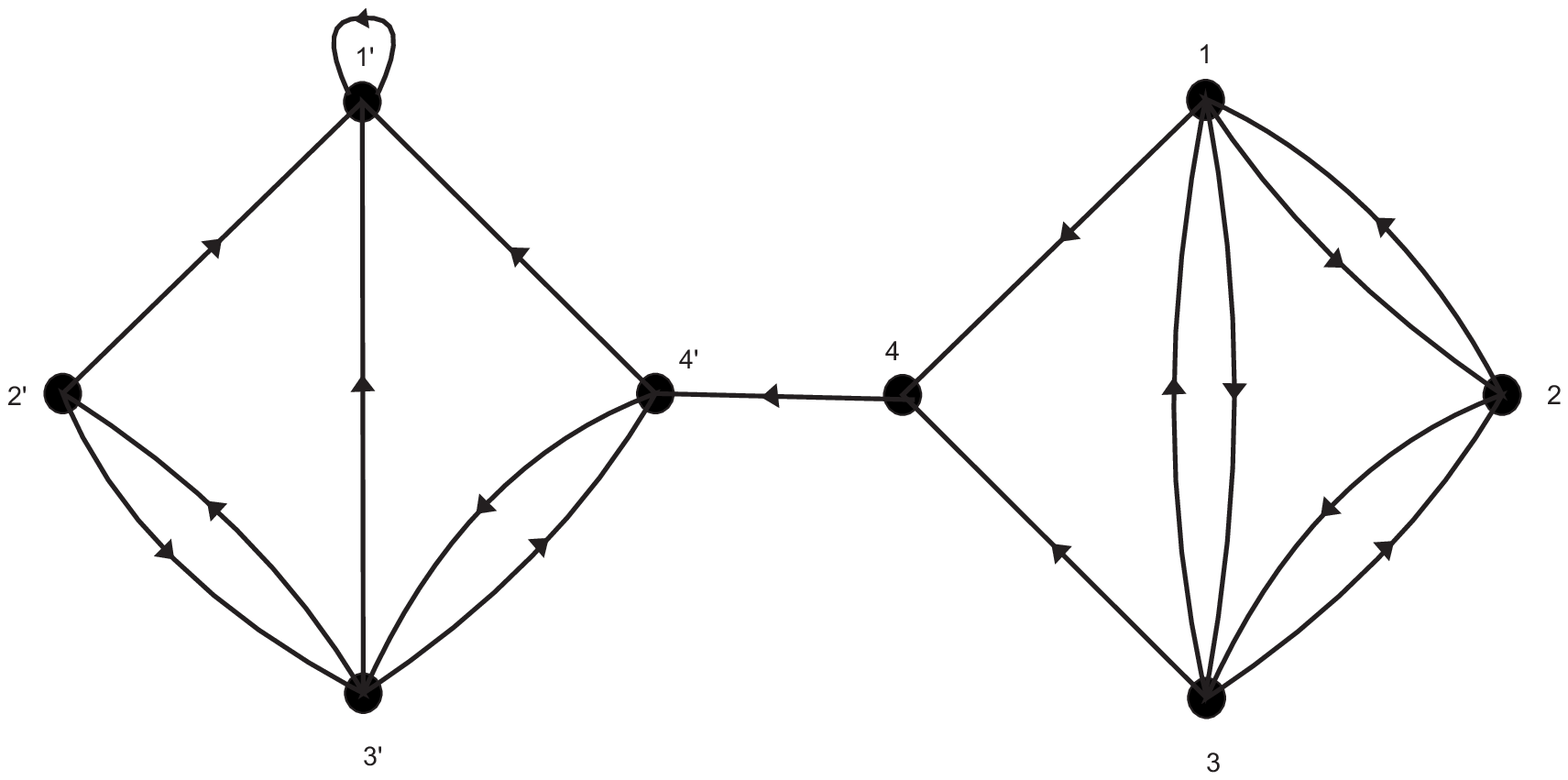}}
\caption{Modified random walk for computing the commute time starting from node 1 to reach for the first time the node 4 and then to reach 1 for the first time starting from 4} \label{sl3}\end{figure}

Let us now compute  the probability starting from node 1 to reach for the first time the node 4 and then to reach 1 for the first time starting from 4 in time $t$. For this, we consider the modified random walk shown on Fig.~\ref{sl3}, with the transition matrix given by: 
$$
C = \left[
\begin{array} {cccccccc}
0 & 1/3 & 1/3 & 1/3 & 0 & 0 & 0 & 0  \\
1/2 & 0 & 1/2 & 0  & 0 & 0 & 0 & 0   \\
1/3 & 1/3 & 0 & 1/3  & 0 & 0 & 0 & 0 \\
0 & 0 & 0 & 0  & 0 & 0 & 0 & 1  \\
0 & 0 & 0 & 0 & 1 & 0 & 0 & 0  \\
0 & 0 & 0 & 0  & 1/2 & 0 & 1/2 & 0   \\
0 & 0 & 0 & 0  & 1/3 & 1/3 & 0 & 1/3 \\
0 & 0 & 0 & 0  & 1/2 & 0 & 1/2 & 0
\end{array}
\right].
$$
The element $c_{15}^t$ of the matrix $C^t$ is the cumulative distribution function of the commute time $\kappa_{14}-1$ and it is given by: 
\begin{eqnarray}
c_{15}^t=  \frac{2^{-t-2} 3^{-t}}{13}&\times& \bigg[13\times 2^t 3^{t/2} \left(3+2 \sqrt{3}+4 \times  3^{t/2}\right)\nonumber\\
&-&\left(1+\sqrt{13}\right)^t \left(65+19 \sqrt{13}\right)\nonumber\\
&+&\left(1-\sqrt{13}\right)^t   
\left(-65+19 \sqrt{13}\right)\nonumber\\
&+&(-2)^t 3^{t/2} \left(39-26 \sqrt{3}\right)\bigg]. 
\end{eqnarray} 
Notice that again  $\lim_{t \to \infty} c_{15}^t = 1$.

\begin{figure}\center
\resizebox{0.5\textwidth}{!} {\includegraphics{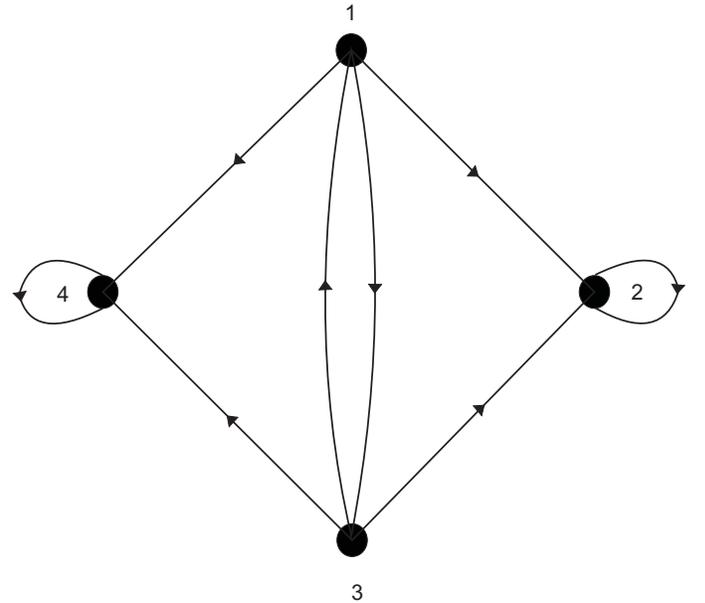}}
\caption{Modified random walk for computing the probability of reaching the node 4 or the node 2 from arbitrary node} \label{sl4}\end{figure}

As the last example, we consider the probability of reaching the node 4 or the node 2 from the node 1 for the first time in time steps $t$. The modified random walk is shown on Fig.~\ref{sl4} and the transition probability matrix of the modified walk is given by the matrix $D_{4;2}$, which has the form: 
$$
D_{4;2} = \left[
\begin{array} {cccc}
0 & 1/3 & 1/3 & 1/3   \\
0 & 1 & 0 & 0     \\
1/3 & 1/3 & 0 & 1/3   \\
0 & 0 & 0 & 1   
\end{array}
\right].
$$
The elements $\tilde{d}_{1,2}^t$ and $\tilde{d}_{1,4}^t$ of the matrix $D_{4;2}^t$ are 
$$
\tilde{d}_{12}^t =\tilde{d}_{14}^t =\frac{1}{2}-\frac{3^{-t}}{2}
$$
The cumulative distribution function of the event: the node 2 or the node 4 is reached from the node 1 for the first time by the  $t$-th step, is given by $\tilde{d}_{12}^{t} + \tilde{d}_{14}^{t}$.

\subsection{Cover time for complete, cycle and path graph} 

In this subsection we derive exact expressions for the CDF of cover time for three particular graphs: complete, cycle, and path graph.

\subsubsection{Complete graph}

A complete graph is a simple graph in which every
pair of distinct vertices is connected by an edge. The
complete graph on $n$ vertices has $n$ vertices and $n(n-1)/2$
edges, and is denoted by $K_n$. 
We can now easily derive analytical results for the PMF of a complete graph. It is easy to see that for the complete graph we have
\begin{eqnarray*}
  P(E_{x_i}^t) &=& 1-\frac{(n-2)^{t}}{(n-1)^t} \\
  P(E_{x_i}^t \cup E_{x_j}^t )&=&1-\frac{(n-3)^{t}}{(n-1)^t} \\
  P(E_{x_i}^t \cup E_{x_j}^t \cup E_{x_k}^t) &=&1-\frac{(n-4)^{t}}{(n-1)^t}. 
\end{eqnarray*}
Therefore, 
\begin{eqnarray*}
  P\bigg(\bigcup_{i=1}^k E_{x_i}^t \bigg)=1-\frac{(n-k-1)^{t}}{(n-1)^t}
\end{eqnarray*}

Thus, the cumulative distribution function of the cover time for complete graph with $n$ nodes can be expressed as
\begin{eqnarray*}
    F_{\mbox{cover}}(t)& = &\sum_{\gamma=1}^{n-1} (-1)^{\gamma-1} \frac{(n-1)!}{\gamma !(n-\gamma-1)!}P\bigg(\bigcup_{i=1}^\gamma E_{x_i}^t \bigg)\nonumber\\
&=&\sum_{\gamma=1}^{n-1} (-1)^{\gamma-1} 
\frac{\Gamma(n)}{\gamma !\Gamma(n-\gamma)}
\bigg(1-\frac{(n-\gamma-1)^{t}}{(n-1)^t}\bigg)\nonumber\\
%&=&\frac{(-1)^n+1}{2} - \sum _{\gamma=1}^{n-1} (-1)^{\gamma-1} \left(1-\frac{\gamma}{n-1}\right)^t\nonumber\\
\end{eqnarray*}
Therefore, the probability mass function is:
\begin{eqnarray*}
f_c(t)=\sum_{\gamma=1}^{n-1}(-1)^{\gamma-1}\frac{\Gamma(n-1)}{\Gamma(\gamma)\Gamma(n-\gamma)}\bigg(1-\frac{\gamma}{n-1}\bigg)^t. 
\end{eqnarray*}

\subsubsection{Cycle graph}

\begin{figure}\center
\resizebox{0.4\textwidth}{!} {\includegraphics{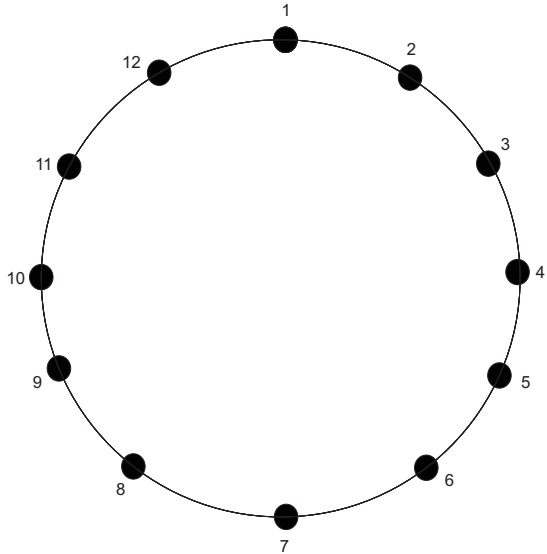}}
\caption{Cycle graph with 12 nodes} \label{fig7}\end{figure}

A cycle graph is a graph that consists of a single cycle,
or in other words, some number of vertices connected in
a closed chain. Let us denote the cycle graph with $n$ vertices as
$C_n$. The number of vertices in a $C_n$ equals the number
of edges, and every vertex has degree 2; that is, every
vertex has exactly two edges incident with it. An example of a cycle graph with 12 nodes is given in Fig.~\ref{fig7}.

Let us assume that the first node of the cycle graph is the starting node of the walk. We need to find the intersection of the events of reaching nodes 2,3, $\ldots$  to $n$. These events form a path. 
A path in a graph is a sequence of vertices such that, from each of its 
vertices, there is an edge to the next vertex in the sequence. 
A cycle is a path such that the start vertex and end vertex are the same. 
Note that the choice of the start vertex in a cycle is arbitrary.
By  exploiting the Remark of Corollary 3.1.16 given in \cite{dohmen}, and proved in \cite{naiman} for events that form a path, we find that the cumulative distribution function of the cover time for a cycle graph is:
\begin{eqnarray*}
    F_{\mbox{cover}}(t)& = &\sum_{i=2}^{n} P\left(E_{i}^t\right)-\sum_{i=2}^{n-1} P\left(E_{i}^t\cup E_{i+1}^ t\right). \quad
\end{eqnarray*}

\subsubsection{Path graph}
\begin{figure}\center
\resizebox{0.4\textwidth}{!} {\includegraphics{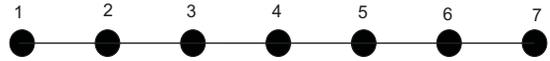}}
\caption{Path graph with 7 nodes} \label{sl_path}\end{figure}

A path graph is a particularly simple example of a tree,
namely one which is not branched at all, that is, contains
only nodes of degree two and one. In particular, two of its
vertices have degree 1 and all others (if any) have degree
2. An example of a path graph with 7 nodes is given in Fig.~\ref{sl_path}.

To find the cumulative  distribution function of the cover time for a path graph we note that all the nodes will be covered if the first and the last nodes are reached by the random walker. Therefore, the cumulative distribution function of the cover time for a path graph is
\begin{eqnarray*}
    F_{\mbox{cover}}(t)& = & P\left(E_{1}^t\cap E_{n}^ t\right)\nonumber\\
&=&P\left(E_{1}^t\right)+P\left( E_{n}^ t\right)-P\left(E_{1}^t\cup E_{n}^ t\right). 
\end{eqnarray*}
We note that if the first node is the starting node then $P\left(E_{1}^t\cap E_{n}^ t\right)=P\left(E_{n}^ t\right)$ and if the last node is the starting node then $P\left(E_{1}^t\cap E_{n}^ t\right)=P\left(E_{1}^ t\right)$.

%\section{Numerical results} \label{sec-numer}
%
%\subsection{Hitting time and commute time} 
%
%Cumulative probability distributions for hitting and commute time can be %computed for  

\section {Approximation of the CDF of cover time} \label{sec-approx} 

The cumulative  distribution functions for hitting and commute time can be computed for reasonable large graphs.  
The complexity of matrix multiplication, if carried out naively, is $O(n^3)$, but more efficient algorithms do exist, which are computationally interesting for matrices with dimensions $n > 100$ \cite{press}.

The inclusion-exclusion formula (\ref{cov-t}) has little practical value in graphs with large number of nodes since it then requires extensive computational times. In the following, we present an accurate and useful approximation of (\ref{cov-t})  that can be evaluated in a reasonable time. The first inequality for the inclusion-exclusion was discovered by Ch. Jordan \cite{jordan} and from then until now a lot of work has been done in sharpening the bounds or the approximation. An excellent survey of the various results for the inclusion-exclusion  is given in  \cite{dohmen}.

We propose the following approximation for inclusion-exclusion formula:
\begin{eqnarray}
    &&P\bigg( \bigcap _{k=1}^n E_{x_k}\bigg)\approx  \prod_{i=1}^n P(E_{x_i})
\nonumber\\
&&\qquad\qquad\times \prod_{i=1}^{n-1}\frac{ P(E_{x_i}\cap E_{x_{i+1}})}{P(E_{x_i}) P(E_{x_{i+1}})}   \label{aprox-1} 
\end{eqnarray}
where $P(E_{x_i}\cap E_{x_{i+1}}) = P(E_{x_i})+P(E_{x_{i+1}})-P(E_{x_i}\cup E_{x_{i+1}})$.
The node indexes must be arranged in such way that there exists an edge between nodes $x_i$ and $x_{i+1}$.  This condition is not strict, and there can exist a small number of nodes that do not satisfy this condition.
The Appendix presents the heuristic derivation of (\ref{aprox-1}) by using the method proposed in \cite{kessler}.

As can be seen, the single-step computational complexity of Eq. (\ref{aprox-1}) is  $O(2 n)$.
The proposed approximation is very accurate for strongly connected graphs like the complete graph and is less accurate for poorly connected graphs like the path graph, as can be seen from the figures below. We note that the error of the approximation for the path graph is the upper bound of the error when the middle node is the starting node. 
This is due to the fact that the proposed approximation equation reduces to the exact equation for independent events, while diminishing as the events become more and more dependent. When almost all the nodes are subset of the rest of the nodes then the approximation formula is the least accurate. Thus the formula is the least accurate when is applied to a path graph with $n$ nodes and the walk starts from the middle, $n/2$-th node (assuming that $n$ is an even number).  In this case the event of reaching nodes 1 to $n/2-1$ is the event of reaching node $1$ and the event of reaching nodes $n/2+1$ to $n$ is the event of reaching node $n$.

\begin{figure}\center
\resizebox{0.5\textwidth}{!} {\includegraphics{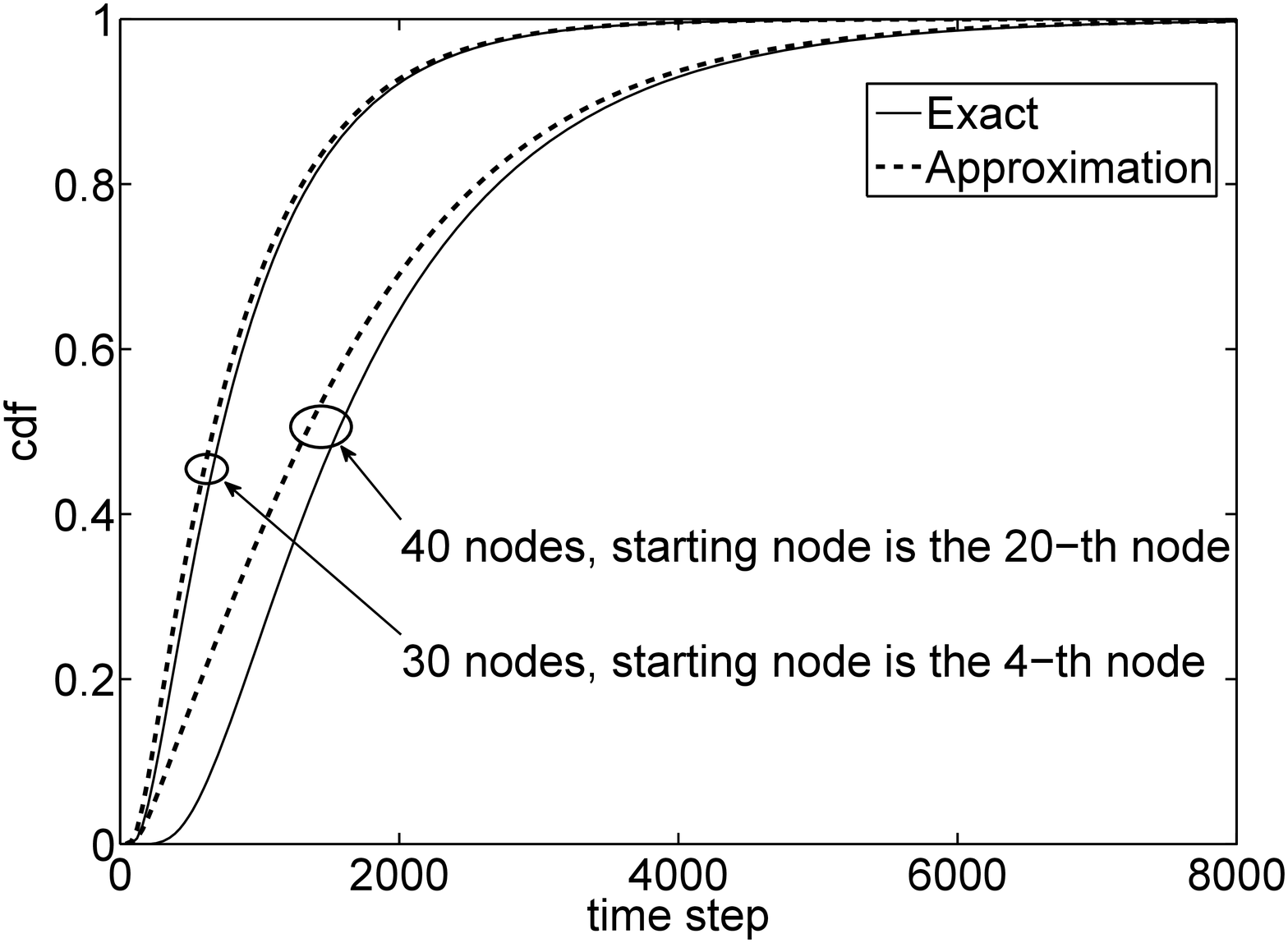}}
\caption{Exact and the approximate formula of the CDF for a path graph with two different starting nodes} \label{sporedba}
\end{figure}

Interestingly, if we start changing the starting node and evaluate the error of the approximation for a path graph, then when the starting node approaches the first or the $n$-th node the formula becomes more and more accurate and when the starting node is the first or the last node, then the approximation formula reduces to the exact formula. To prove this we let the first node be the starting node, then  the event of reaching  node  $i$ and node $j$, if $j>i$ is $P( E_i^t\cap E_j^t)=P(E_j^t)$. Thus the approximation formula is given by
\begin{eqnarray} \label{dependent}
 P\bigg( \bigcap _{k=1}^n E_{k}\bigg)=  \prod_{i=1}^n P(E_{i})
 \prod_{i=1}^{n-1}\frac{1}{P(E_{i})}=P(E_n)
\end{eqnarray}
A similar proof is when the $n$-th node is the starting node.
An example is given in Fig.\ref{sporedba} where the CDF of a path graph is given by the exact and the approximate formula for two path graphs with 30 and 40 nodes when starting node is the 4-th and the 20-th node, respectively. The second worst case error is when the approximation formula is applied to a cycle graph and in this case the error is independent of the starting node.

 \begin{figure}\center
(a) 
\resizebox{0.5\textwidth}{!} {\includegraphics{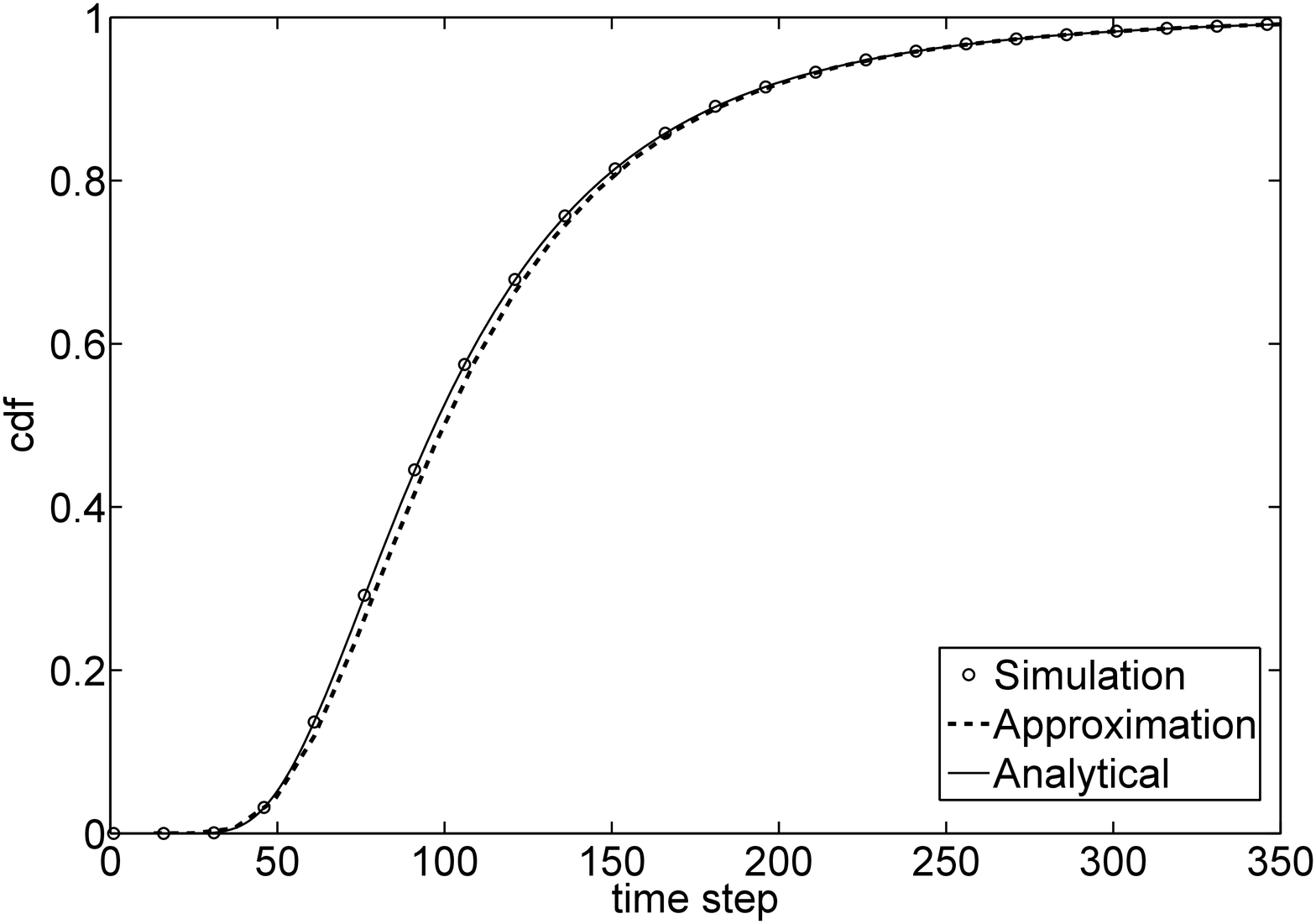}} \\
(b) 
\resizebox{0.5\textwidth}{!} {\includegraphics{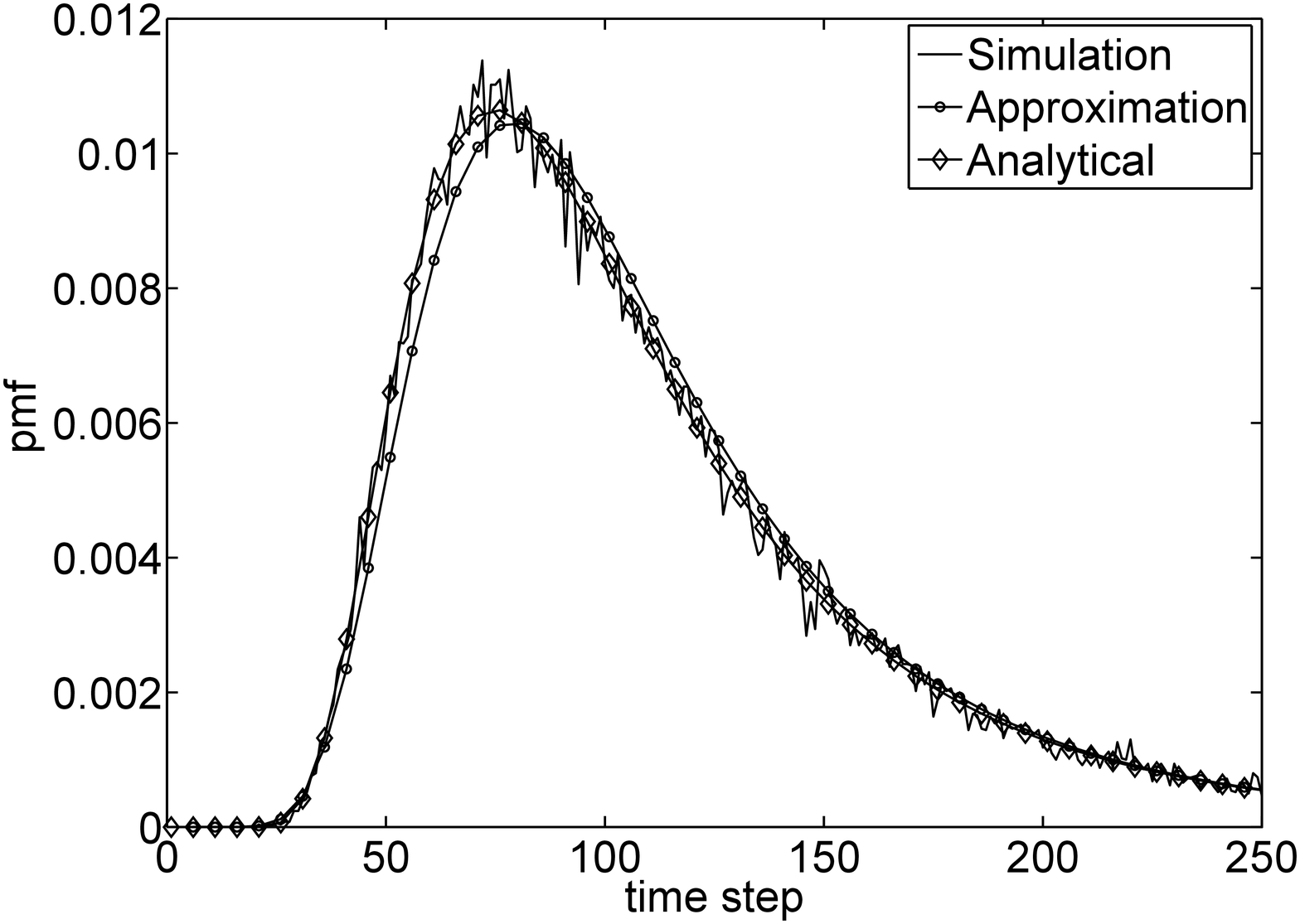}}
\caption{Analytical, approximated and simulated (a) cumulative distribution function  and (b) probability mass function of cover time for a random graph with 20 nodes} \label{fig1}\end{figure}

\begin{figure}\center
(a) 
\resizebox{0.5\textwidth}{!} {\includegraphics{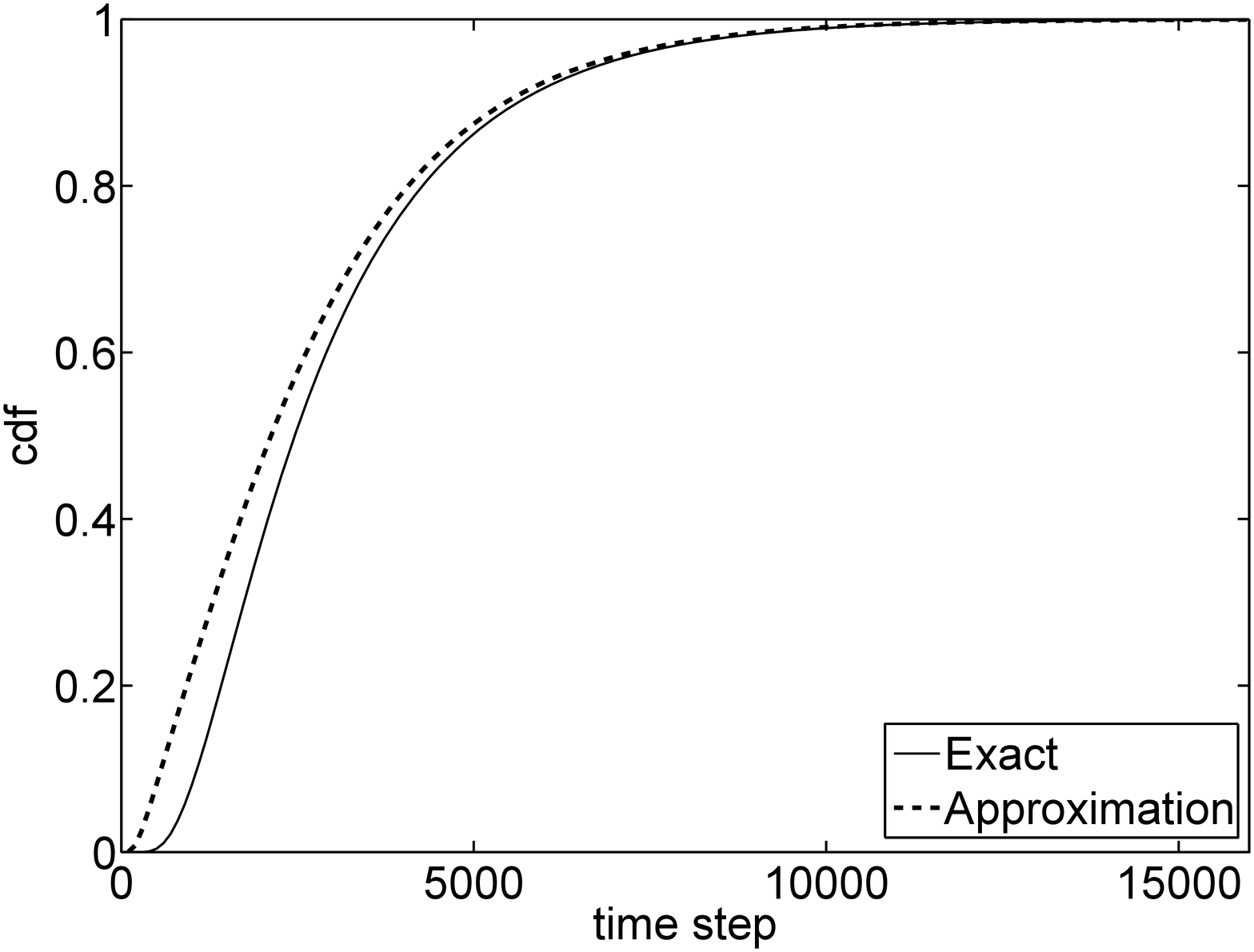}} \\
(b) 
\resizebox{0.5\textwidth}{!} {\includegraphics{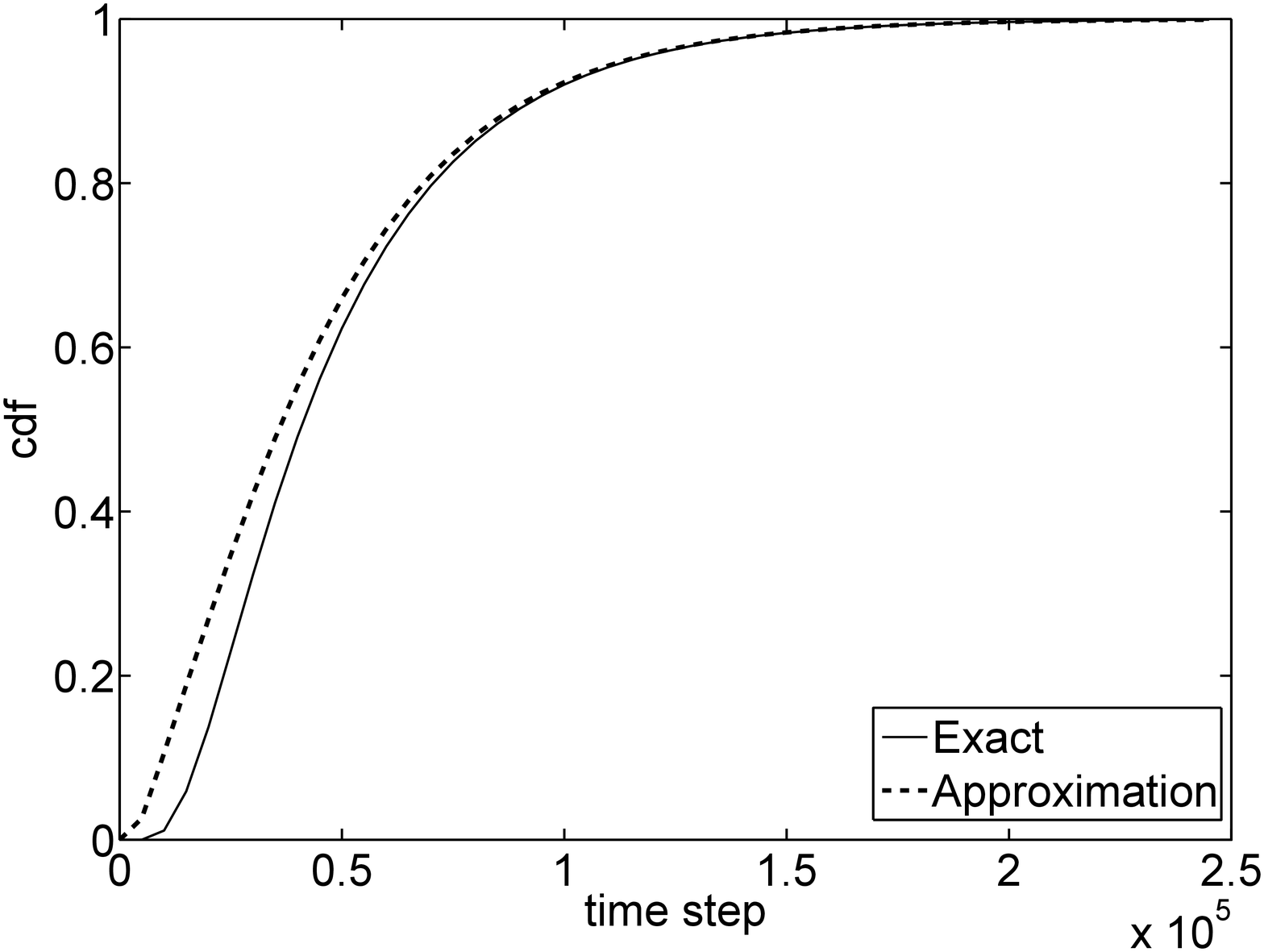}}
\caption{Exact vs approximated CDF for (a) a path graph with 50 nodes and (b) a path graph with 200 nodes } \label{fig22}\end{figure}

\begin{figure}\center
(a) 
\resizebox{0.5\textwidth}{!} {\includegraphics{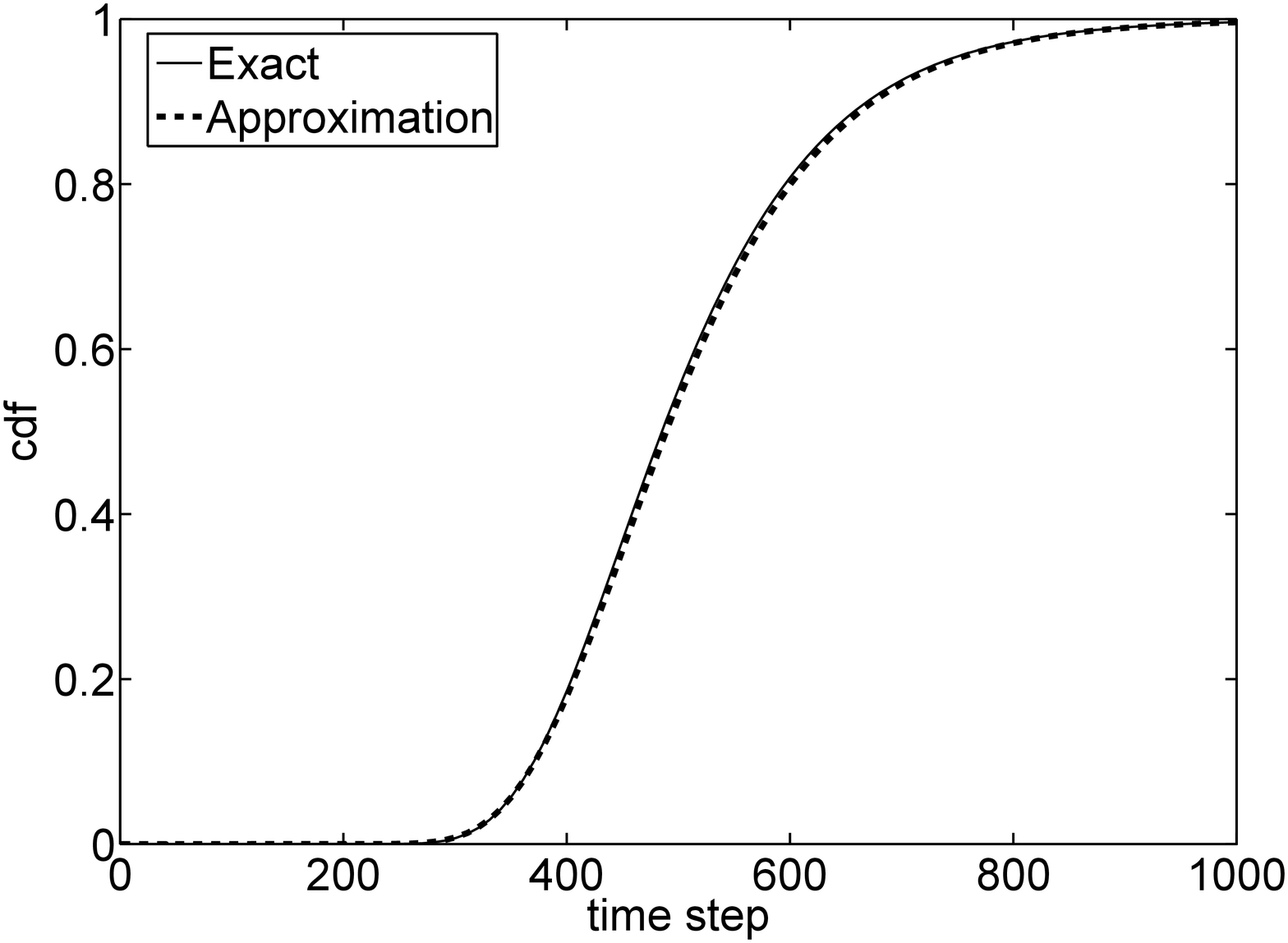}} \\
(b) 
\resizebox{0.5\textwidth}{!} {\includegraphics{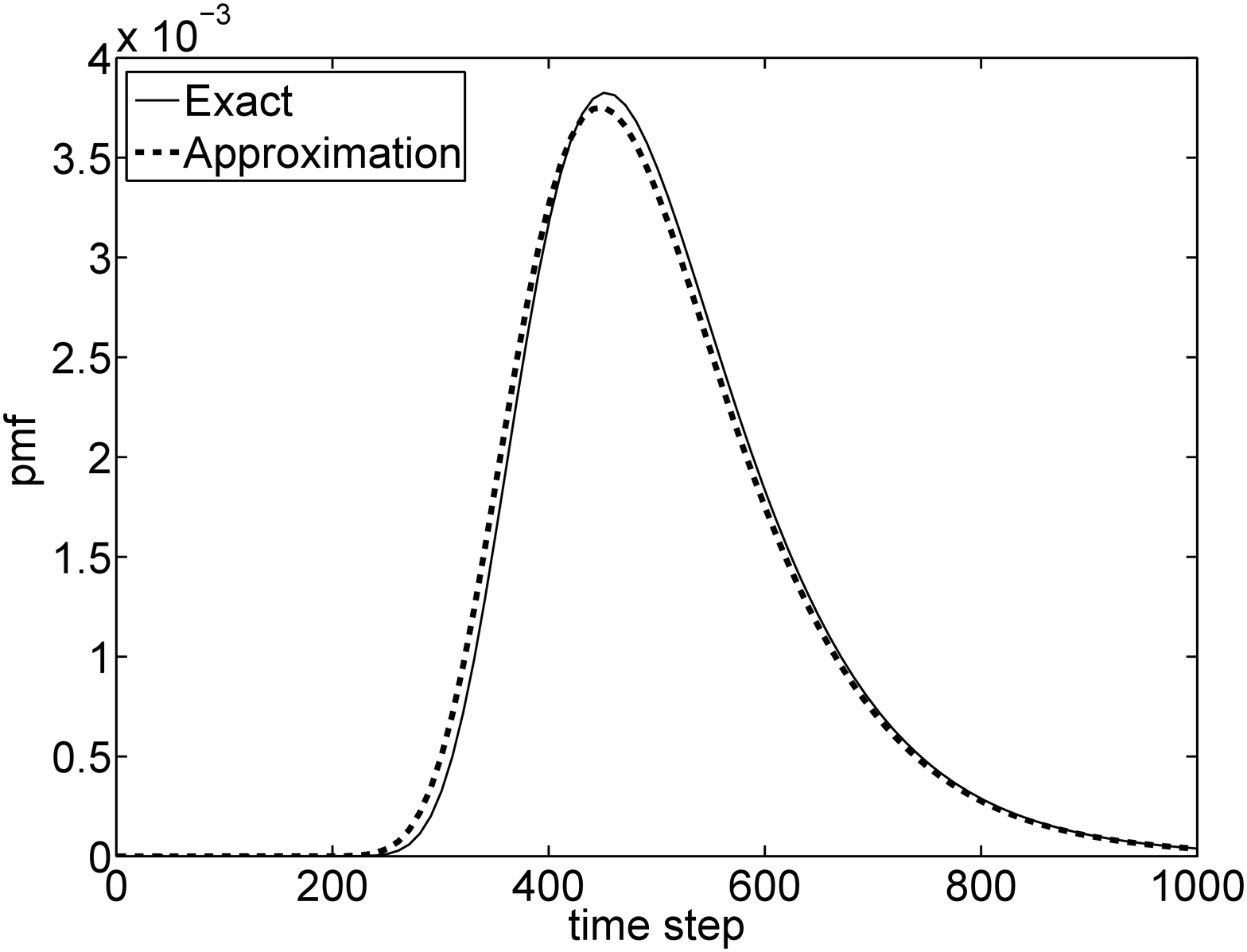}}
\caption{Exact vs approximated CDF (a) and PMF (b) for complete graph with 50 nodes } \label{fig3}\end{figure}

\section{Numerical Examples} \label{sec-num}

In this section, several numerical examples are presented. 
First, we validate the cover time formula and the approximation by Monte Carlo simulations, Fig.~\ref{fig1}, for a Erdos-Renyi random graph with 20 nodes. Fig.~\ref{fig1}a illustrates the CDF, while Fig.~\ref{fig1}b the PDF of cover time.
We illustrate the accuracy of the approximation for a path graph and a complete graph, Figures \ref{fig22} and \ref{fig3} respectively, where the  starting node of the walk for the path graph is the middle node for both Figures \ref{fig22}a and \ref{fig22}b.     
We have performed various numerical simulations of the cumulative distribution functions using exact and approximate expressions for complete, path and cycle graphs with up to 1000 nodes. For small $n$ ($n \leq 1000$) we found that increasing $n$ to up to 1000, the accuracy of the approximation is maintained.   We believe that  Eq.~(\ref{aprox-1}) is a good approximation for cumulative distribution of cover time even for larger graphs, but since at the moment we do not have estimates for accuracy of our approximation, we leave this as a subject of our next research. More detailed analysis on how the CDF of cover time depends on graph topology will be discussed in a forthcoming paper.

%\begin{figure}\center
%\resizebox{0.5\textwidth}{!} {\includegraphics{path_sporedba.eps}}
%\caption{CDF of the Erdos-Renyi random graph with 500 nodes} %\label{fig-cdf-ER-500}
%\end{figure}

%\subsection{Cover time for different topologies} 

\section{Conclusions} 

In this paper we have derived the exact closed-form expressions for the PMF and CDF of three random walk parameters  that play pivotal role in the theory of random walks: hitting time, commute time, and cover time.  We also have derived simpler closed formulas for the  cumulative distribution function of cover time for complete, cycle and path graphs. An approximation of the  cumulative distribution function for cover time is proposed, and several numerical results for the CDF of cover time for different graphs are presented.

\section*{Appendix}   

If A is the union of the events $A_1, A_2, \ldots,  A_n$ then, writing $p_i$ for the probability of $A_i$, $p_{ij}$ for the probability of $A_i \cap  A_j$ , $p_{ijk}$ for the probability of $A_i \cap A_j \cap A_k$ etc, the probability of $A$ is given by
$$
p(A)= \sum_i p_i - \sum_{i<j} p_{ij} + \sum_{i<j<k}  p_{ijk} - \ldots + (-1)^{n-1} p_{12\ldots n}   
$$
The inclusion -- exclusion principle tells us that if we know the $p_i, p_{ij}, p_{ijk} \ldots $ then we can find $P(A)$. However, in practice we are unlikely to have full information on the $p_i, p_{ij}, p_{ijk} \ldots $. Therefore,  
we are faced with the task of approximating $P(A)$ taking into account
whatever partial information we are given. 
In certain cases where the events $A_i$ are in some sense close to being independent,
then there are a number of known results approximating $P(A)$. In this paper we use the following result [\cite{kessler}, equation (9)]: 
\begin{equation}\label{app}
P(A) \approx 1 - \prod_{i=1}^n ( q_i) \prod_{i=1}^n \prod_{j=i+1}^n (q_{ij} )
\end{equation}
where 
\begin{eqnarray}
q_i&=&P(\bar A_i)\\
q_{ij}&=&\frac{P(\bar A_i\cap \bar A_j)}{P(\bar A_i)P(\bar A_j)}
\end{eqnarray}
Let the event $B_i$ be defined as $B_i=\bar A_i$. Then   $B=\bigcup_{i=1}^n  B_i=\bigcup_{i=1}^n \bar A_i$ and the probability of this event is:
\begin{equation}\label{app1}
P(B)=P\left(\bigcup_{i=1}^n \bar A_i\right)=1-P\left(\bigcap_{i=1}^n  A_i\right)
\end{equation}
The approximated form (\ref{app}) of the event $B$ is:
\begin{equation}\label{app2}
P(B) \approx 1 - \prod_{i=1}^n P(\bar B_i) \prod_{i=1}^n \prod_{j=i+1}^n \frac{P(\bar B_i\cap \bar B_j)}{P(\bar B_i)P(\bar B_j)}
\end{equation}
Replacing (\ref{app1}) and $\bar B_i=A_i$ into the (\ref{app2}), we get:
\begin{eqnarray}\label{app_final}
 P\left(\bigcap_{i=1}^n  A_i\right)\approx \prod_{i=1}^n  P( A_i) \prod_{i=1}^n \prod_{j=i+1}^n\frac{P( A_i\cap A_j)}{P( A_i)P( A_j)},
\end{eqnarray}
where $P( A_i\cap A_j)$ can be expressed as:
$$
P( A_i\cap A_j)=P(A_i)+P(A_j)-P(A_i\cup A_j)
$$
When the events $A_i$ and $A_j$ are not close to being independent but on the contrary, one of the events  is a subset of the other, as the case for the events in the cover time formula, the approximation formula  (\ref{app_final}) is not accurate.

The inaccuracy can be seen from the following example: Let the events $A_{j}$ for $j>i$ are all subsets of the event $A_i$. Then the probability of the event $ A_i\cap A_j$ is
$$
P( A_i\cap A_j)=P(A_j)
$$
if we now replace this expression in (\ref{app_final}) we get
$$
P\left(\bigcap_{i=1}^n  A_i\right)\approx \prod_{i=1}^n  P( A_i) \prod_{i=1}^n\left(\frac{1}{P( A_i)}\right)^{n-i}
$$
Then if $n$ is large and the probabilities $P(A_i)$ are a very small numbers, this probability expression can be a number much bigger then one. 

One way to solve the accuracy problem is not to take  the second product  over all node pairs, but just over $n-1$ different neighboring pairs. We suggest the following approximation for the cumulative distribution of cover time: 
\begin{eqnarray*}\label{app_final-1}
 P\left(\bigcap_{i=1}^n  A_i\right)\approx \prod_{i=1}^n  P( A_i) \prod_{i=1}^{n-1}\frac{P( A_i\cap A_{i+1})}{P( A_i)P( A_{i+1})}.
\end{eqnarray*}
We note that this approximate probability expression reduces to the exact probability expression in the two limiting cases: first, when all events are mutually independent, and second, when all events are subset of just one event.
The first claim can be proved just by noting that  $P( A_i\cap A_{i+1})=P( A_i)P( A_{i+1})$ and the second claim was previously proved, see equation (\ref{dependent}) when the events $E_i$ for $i=1, \ldots, n-1$ are all subsets of the event $E_n$.

\end{document}